\documentclass{article}
\usepackage{ijcai24}

\usepackage{times}
\usepackage{soul}
\usepackage{url}
\usepackage[hidelinks]{hyperref}
\usepackage[utf8]{inputenc}
\usepackage[small]{caption}
\usepackage{graphicx}
\usepackage{amsmath}
\usepackage{amsthm}
\usepackage{booktabs}
\usepackage{algorithm}
\usepackage{algorithmic}
\usepackage[switch]{lineno}

\usepackage{subfigure}
\usepackage{amssymb}
\usepackage{diagbox}
\usepackage{makecell}

\title{JointPPO: Diving Deeper into the Effectiveness of PPO in Multi-Agent Reinforcement Learning}

\author{
Chenxing Liu\And
Guizhong Liu\\
\affiliations
School of Information and Communications Engineering, Xi'an Jiaotong University\\
\emails
lcx459455791@stu.xjtu.edu.cn,
liugz@xjtu.edu.cn
}

\begin{document}

\maketitle

\begin{abstract}
While \textit{Centralized Training with Decentralized Execution} (CTDE) has become the prevailing paradigm in Multi-Agent Reinforcement Learning (MARL), it may not be suitable for scenarios in which agents can fully communicate and share observations with each other. Fully centralized methods, also know as \textit{Centralized Training with Centralized Execution} (CTCE) methods, can fully utilize observations of all the agents by treating the entire system as a single agent. However, traditional CTCE methods suffer from scalability issues due to the exponential growth of the joint action space. To address these challenges, in this paper we propose JointPPO, a CTCE method that uses Proximal Policy Optimization (PPO) to directly optimize the joint policy of the multi-agent system. JointPPO decomposes the joint policy into conditional probabilities, transforming the decision-making process into a sequence generation task. A Transformer-based joint policy network is constructed, trained with a PPO loss tailored for the joint policy. JointPPO effectively handles a large joint action space and extends PPO to multi-agent setting in a clear and concise manner. Extensive experiments on the StarCraft Multi-Agent Challenge (SMAC) testbed demonstrate the superiority of JointPPO over strong baselines. Ablation experiments and analyses are conducted to explores the factors influencing JointPPO's performance.
\end{abstract}


\section{Introduction}
Multi-Agent Systems (MAS) are complex systems composed of multiple agents that cooperate with each other interacting within a common environment~\cite{mas}. Such systems are ubiquitous in various real-world scenarios, including traffic light control~\cite{traffic_light}, finance~\cite{finance}, and robots coordination~\cite{robot_coordination}. In this paper, we aim at developing an intelligent decision-making method for fully cooperative MAS, in which agents act as a unified team to tackle complex tasks.

While Reinforcement Learning (RL) has shown remarkable success in achieving intelligent decision-making and control~\cite{muzero,daydreamer}, applying RL to multi-agent systems, known as Multi-Agent Reinforcement Learning (MARL), emerges as a promising and challenging research area. In MARL, all the agents consistently interact with the environment, optimizing their policies in a trial-and-error manner, with the goal of maximizing the expected cumulative rewards.

\begin{figure*}
\centering
\subfigure[Fully independent (DTDE) paradigm.]{\label{fig1a}
\includegraphics[width=0.3\linewidth]{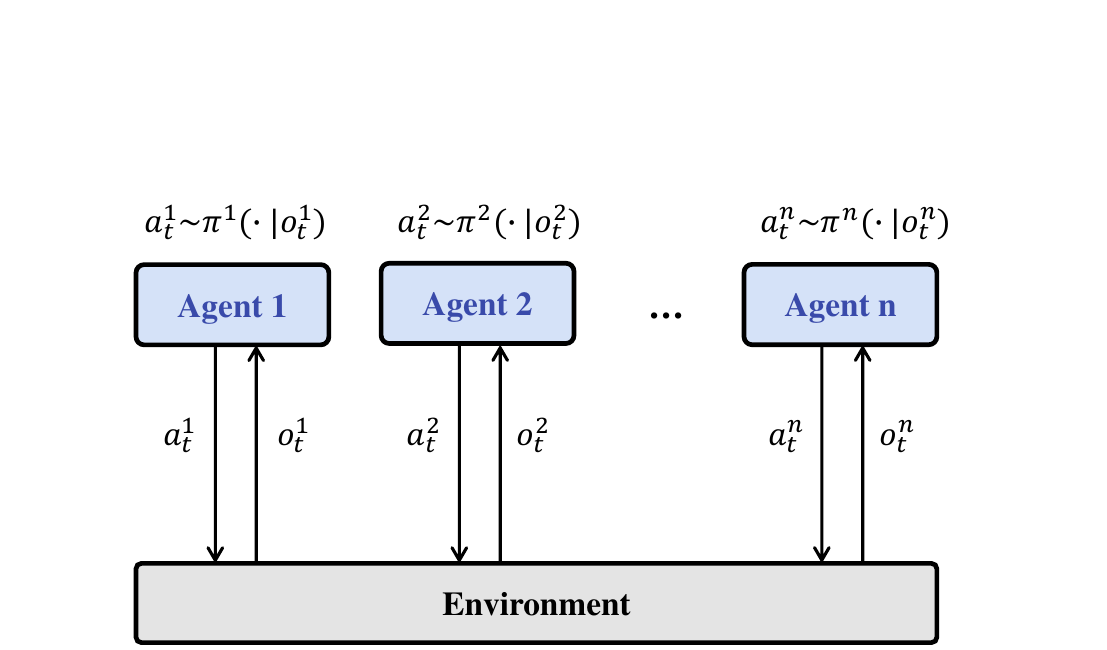}}\hspace{-3mm}
\subfigure[Fully centralized (CTCE) paradigm.]{\label{fig1b}
\includegraphics[width=0.3\linewidth]{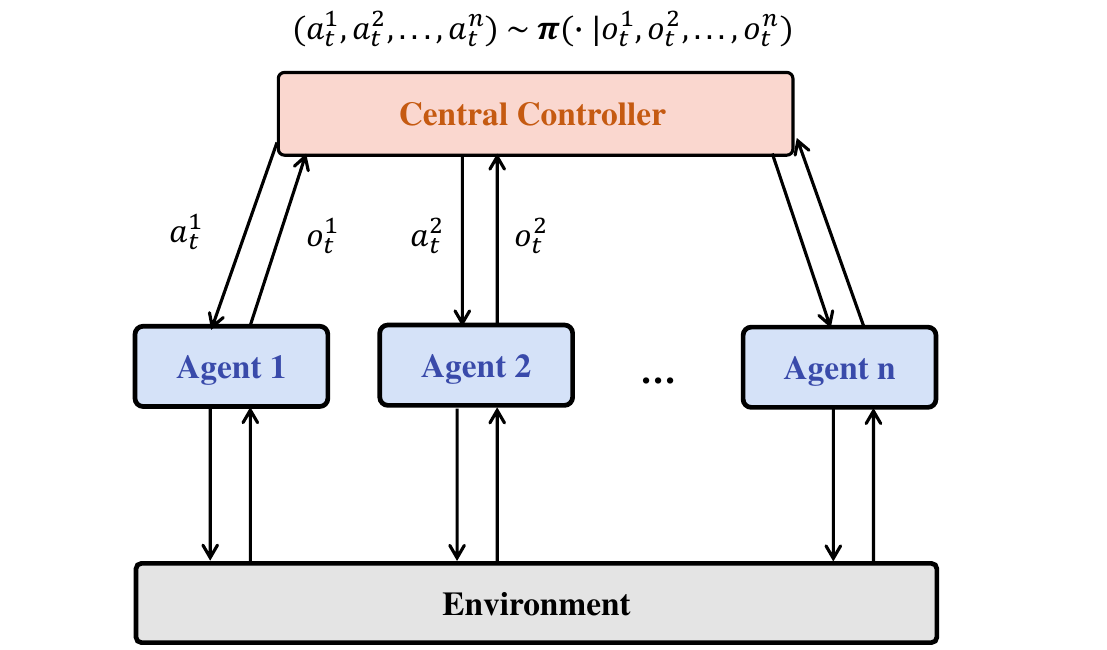}}
\subfigure[CTDE paradigm.]{\label{fig1c}
\includegraphics[width=0.3\linewidth]{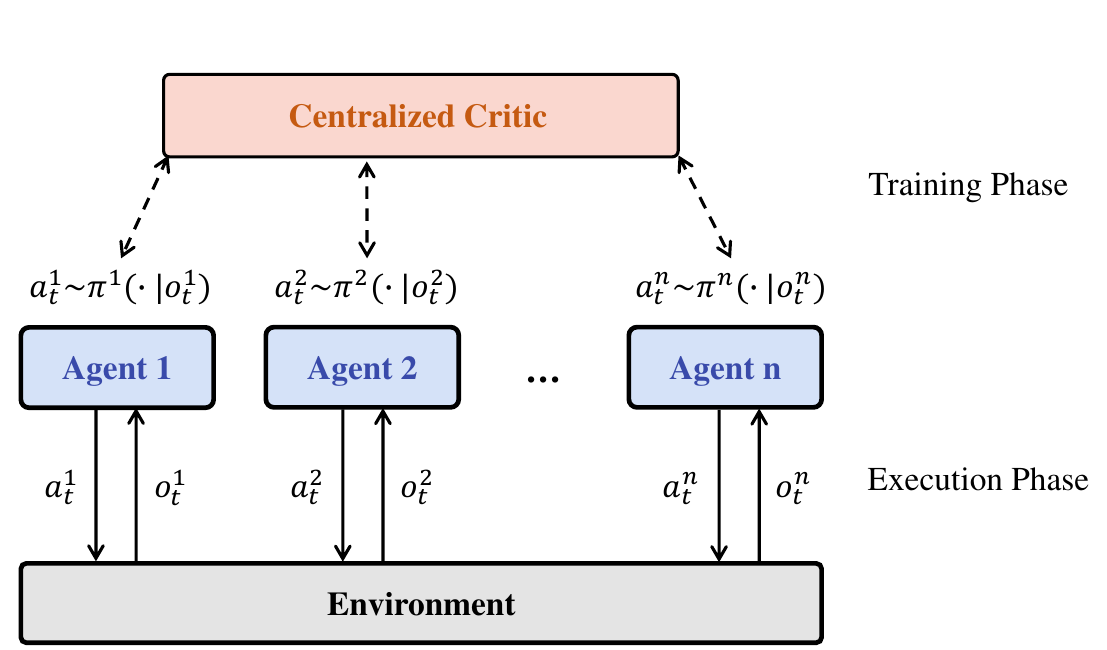}}
\caption{Different learning paradigms in MARL.}
\label{fig1}
\end{figure*}

The existence of more than one learning agent in MARL incurs greater uncertainty and learning instability, making it a challenging problem~\cite{review_of_challenge}. To tackle this challenge, existing MARL algorithms primarily fall into three categories: fully independent methods, fully centralized methods, and \textit{centralized training with decentralized execution} (CTDE) methods. In fully independent methods, as depicted in Figure \ref{fig1a}, each agent trains and acts independently, making decisions solely based on its own observation~\cite{ippo,iql}. This allows for direct implementation of single-agent RL algorithms, but still suffers from learning instability. CTDE methods, on the other hand, allow agents to access the environment state or all agents' observations during training while still maintaining decentralized execution during interactions~\cite{coma,maddpg,qmix}, which can be depicted as Figure \ref{fig1c}. While both of these methods have achieved good performance, they are limited to decentralized execution, which means that agents share little information with each other and make decision independently. Such limitation makes them less appropriate for scenarios with sufficient communication, as the information available in other agents' observations is ignored. Additionally, the widespread use of parameter-sharing techniques in these methods leads to suboptimal outcome and homogeneous behavior, potentially causing failure in complex scenarios~\cite{happo}.

Alternatively, fully centralized methods, also known as \textit{centralized training with centralized execution} (CTCE) methods, mitigate these limitations by treating the entire multi-agent system as a single agent and making full use of all the agents' observations~\cite{csrl,copa}. Figure\ref{fig1b} illustrates CTCE methods that behave as a central controller which integrates all agents' observation and designates their actions. Current centralized methods, however, face challenges due to the exponential growth of joint action spaces with increasing number of agents. 

In this paper, we address those challenges in two steps: First, we decompose the joint policy of the multi-agent system into conditional probabilities, transforming the decision-making process into a sequence generation task, and propose a general framework that solves MARL with any sequence generation model. The proposed framework consists of three modules: a Decision Order Designation Module, a Joint Policy Network, and a Centralized Critic Network. These modules are connected in a proper manner to facilitate an effective learning and inference pipeline within the sequence generation scheme. 

Then we propose JointPPO, a CTCE method designed to directly optimize the joint policy of the multi-agent system. As an instance of the proposed framework, JointPPO contains a joint policy network which acts as a central controller, taking all the agents' observations as input and generating agents' actions sequentially at each decision time. Considered that the network architecture design is not the primary focus of this paper, we adopt the Transformer structure introduced in the work of~\cite{mat} with some modifications as our joint policy network. Subsequently, a PPO loss tailored for the joint policy is designed for the network training. As for the Decision Order Designation Module, two choices of decision order designation mechanisms respectively based on prior knowledge and graph generative model are implemented and studied. All these efforts enable JointPPO to achieve a direct optimization of the joint policy without the needs for any value decomposition. As a result, it simplifies MARL to single-agent RL and effectively brings the advantages of PPO to MARL in a clear and concise manner.

We extensively evaluate JointPPO on StarCraft Multi-Agent Challenge (SMAC) testbed~\cite{smac} across various maps, encompassing both homogeneous and heterogeneous scenarios. JointPPO consistently demonstrates superior performance and data efficiency compared to strong baselines. It achieves nearly 100\% win rates across all the test maps and exhibits an remarkable advantage in terms of cost for victory such as killed allies. Comprehensive ablation experiments and analyses are further conducted to explore the elements influencing JointPPO's training process and final performance. Particularly, ablation studies on decision order designation mechanism demonstrate JointPPO's robustness to the order of action generation. 

To sum up, the contributions of this work are as follows:
\begin{itemize}
    \item We explicitly decompose the joint policy of the multi-agent system into conditional probabilities, and propose a general framework of solving MARL using any sequence generation model.
    \item We present JointPPO as an instance of the proposed framework. JointPPO achieves a direct optimization of the joint policy, and effectively handles the high-dimensional joint action space by leveraging the Transformer model.
    \item As a CTCE method, JointPPO's performance demonstrates the feasibility of addressing MARL by simplifying it into single-agent RL, which brings the promising prospective of integrating research outcomes from single-agent RL into the domain of MARL.
\end{itemize}


\section{Related Works}
In recent years, there has been significant progress in MARL. Fully independent method IQL~\cite{iql} first explored extending RL to multi-agent setting by applying DQN to each agent independently. Fully centralized methods, on the other hand, have received less attention since they suffer from scalability issues due to the exponential growth of joint action space. Existing fully centralized methods usually adopt an information exchange mechanism to handle the large action space~\cite{csrl,copa}. However, in practice they do not exhibit the expected stronger performance than CTDE methods.

The CTDE paradigm reaches a compromise between fully independent and centralized approaches, attracting great attention of the MARL community~\cite{ctde2,ctde1}. Numerous works have emerged within the CTDE paradigm, encompassing both value factorization methods and policy gradient methods. Most value factorization methods were designed to satisfy the IGM (Indicidual-Global-Max) condition~\cite{qtran}. VDN~\cite{vdn} first conducted value factorization by approximating the joint value function as a sum of individual ones. QMIX~\cite{qmix} extended this with a monotonicity assumption and a non-negative-weighted mixer network. Subsequent efforts usually built upon the structure introduced in QMIX, further approaching the IGM condition or introducing additional components~\cite{maven,wqmix,qtran,qplex}. However, these value factorization methods face a common challenge caused by the mismatch between the optimal joint value function and individual value functions during training. Such mismatch necessitates more iterations to recover the satisfaction of IGM, resulting in low sample efficiency. 

Among the various policy gradient methods, trust region methods, represented by Trust Region Policy Optimization (TRPO)~\cite{trpo} and Proximal Policy Optimization (PPO)~\cite{ppo}, stand out for their supreme performance with theoretically-justified monotonic policy improvement~\cite{theoretical_guarantee}. Numerous studies have tried to extend this advantage to the multi-agent setting. While IPPO~\cite{ippo} applied PPO independently to each agent, MAPPO~\cite{mappo} introduced centralized critics and comprehensively explored factors that influences its performance. HAPPO~\cite{happo} presented an \textit{Multi-Agent Advantage Decomposition Theorem} and proposed a sequential update scheme. MAT~\cite{mat}, the most relevant work to this paper, was derived from the \textit{Multi-Agent Advantage Decomposition Theorem} and introduced a novel approach of leveraging the sequence model to generate agents' actions sequentially. There are also lots of recent works following the sequential update scheme or the action-dependent scheme~\cite{rpisa,mirror_learning,ace,mamt,a2po,stackelberg}. However, these methods often require intricate theoretical analyses under specific assumptions, which may increase the complexity of practical implementation. In contrast, in this paper we start off from a fully centralized perspective, employ PPO to directly optimize the joint policy of the multi-agent system, thereby achieving a smooth extension of PPO to the multi-agent setting.


\section{Preliminaries}

\subsection{PODMP}
We consider the decision-making problem in the fully cooperative multi-agent systems described by \textit{Partially Observable Markov Decision Processes} (POMDP)~\cite{pomdp}. An $n$-agent POMDP can be formalized as a tuple $\left\langle\mathcal{N,S},\boldsymbol{\mathcal{O,A}}, P, R, \gamma\right\rangle$, where $\mathcal{N}= \left\{1,\dots,n \right\}$ is the set of agents and $\mathcal{S}$ is the global state space of the environment. We denote the local observation and action space of agent $i$ by $\mathcal{O}^i$ and $\mathcal{A}^i$ respectively, and subsequently, the Cartesian product $\boldsymbol{\mathcal{O}}=\mathcal{O}^1 \times, \dots, \times\mathcal{O}^n$ represents the joint observation space of the system while $\boldsymbol{\mathcal{A}}=\mathcal{A}^1 \times, \dots, \times\mathcal{A}^n$ represents the joint action space. $P:\mathcal{S}\times\boldsymbol{\mathcal{A}}\times\mathcal{S}\rightarrow\left[0,1\right]$ is the transition function denoting the state transition probability. $R:\mathcal{S}\times\boldsymbol{\mathcal{A}}\rightarrow\mathbb{R}$ represents the reward function which gives rise to the instantaneous reward and $\gamma\in[0,1)$ is the discount factor that gives smaller weights to future rewards. 

In POMDP, each agent $i\in\mathcal{N}$ has access to only the observation $o^i_t\in\mathcal{O}^i$ to the environment rather than the global state $s_t\in\mathcal{S}$. At each time step $t$, all agents $i\in\mathcal{N}$ choose their actions $a^i_t\in\mathcal{A}^i$, which may be discrete or continuous. Together, these actions form the joint action $\boldsymbol{a}_t=\left(a^1_t,\dots,a^n_t\right)\in\boldsymbol{\mathcal{A}}$. Executing the joint action $\boldsymbol{a}_t$, the agents stimulate the environment into the next state according to the transition function $P$ and, at the same time, receive a scalar team reward $r_t=R\left(s_t, \boldsymbol{a}_t\right)$. Repeating the above process, agents consistently interact with the environment and collect rewards. We define the joint policy $\boldsymbol{\pi}\left(\boldsymbol{a}_t|\boldsymbol{o}_t\right)$ as a conditional probability of the joint action $\boldsymbol{a}_t$ given all the agents' observations $\boldsymbol{o}_t=\left(o^1_t,\dots,o^n_t\right)\in\boldsymbol{\mathcal{O}}$, and return $G\left(\boldsymbol{\tau}\right)=\sum_{t=0}^{\infty}\gamma^tr_t$ as the accumulated discounted rewards, where $\boldsymbol{\tau}$ denotes the sampled trajectory. The goal of MARL is to learn an optimal joint policy $\boldsymbol{\pi}^*$ that maximizes the expected return:
\begin{equation}
\begin{aligned}
\boldsymbol{\pi}^*&=\mathop{\arg\max}_{\boldsymbol{\pi}}\mathbb{E}_{\boldsymbol{\pi}}\left[G\left(\boldsymbol{\tau}\right)\right] \\
&=\mathop{\arg\max}_{\boldsymbol{\pi}}\mathbb{E}_{\boldsymbol{\pi}}\left[\sum_{t=0}^{\infty}\gamma^tR\left(s_t, \boldsymbol{a}_t|_{\boldsymbol{a}_t\sim\boldsymbol{\pi}\left(\boldsymbol{a}_t|\boldsymbol{\boldsymbol{o}_t}\right)}\right)\right].\label{eq1}
\end{aligned}  
\end{equation}

\subsection{Multi-Agent Transformer}

The state-of-the-art algorithm, Multi-Agent Transformer (MAT)\cite{mat}, first effectively solve MARL problem by transforming it into a sequence generation problem and leveraging the Transformer model to map the input of the agents' observations $\left(o^1_t,\dots, o^n_t\right)$ to the output of the agents' actions $\left(a^1_t,\dots,a^n_t\right)$. MAT's training loss is designed based on the \textit{Multi-Agent Advantage Decomposition Theorem}, which decomposes the joint advantage function of the multi-agent system into individual advantage functions and implies a sequential update scheme. However, we argue that the loss in MAT’s implementation does not strictly adhere to the mentioned theorem. To ensure a convergence to the Nash Equilibrium, the theorem suggests that each agent updates its policy towards the best response to the preceding agents' actions, which can be achieved through the policy update described by individual advantage function $A^{i_m}_{\boldsymbol{\pi}}(\boldsymbol{o},\boldsymbol{a}^{i_{1:m-1}},a^{i_m})$. While in MAT, each agent's policy is updated according to the joint advantage function $\boldsymbol{A}^{i_{1:n}}_{\boldsymbol{\pi}}(\boldsymbol{o},\boldsymbol{a}^{i_{1:n}})$, which measures the value of the joint action rather than the value of each agent's action given the preceding agents' actions. We believe this results in some bias in the optimization target. 

Despite those concerns, MAT's use of the Transformer has been a significant contribution and success. Therefore, in this paper, we adopt its Transformer architecture with some modifications to construct our joint policy network. 


\section{Method}
In this section, we present details of JointPPO in four subsections: problem modeling, Transformer-based joint policy network, joint PPO loss, and decision order designation mechanism. In problem modeling, we discuss the decomposition of the joint policy and propose a general framework that solves MARL with sequence generation model. In the next three subsections, we present the detail of JointPPO, as an instance of the proposed framework.
 
\subsection{Problem Modeling}
As mentioned earlier, the goal of MARL is to learn an optimal joint policy $\boldsymbol{\pi}^*\left(\boldsymbol{a}_t|\boldsymbol{o}_t\right)$ that maximizes the expected accumulated return (Eq.~\eqref{eq1}). Most existing methods handle the joint policy $\boldsymbol{\pi}$ by decomposing it into independent individual policies:
\begin{equation}
    \boldsymbol{\pi}\left(\boldsymbol{a}_t|\boldsymbol{o}_t\right)= \boldsymbol{\pi}\left(a^1_t, a^2_t,\dots,a^n_t|\boldsymbol{o}_t\right)=\prod_{i=1}^n\pi^i\left(a^i_t|\boldsymbol{o}_t\right).\label{eq2}
\end{equation}
Notably, there are several different formulations of the individual policy, such as $\pi^i\left(a^i_t|o^i_t\right)$ or $\pi^i\left(a^i_t|\tau^i_t\right)$, which differ in the available information used for decision making. However, the underlying assumption remains unchanged: the agents' actions are independent of each other. 

While such independence assumption facilitates decentralized execution, it has some shortcomings. First, as highlighted in the introduction, decentralized execution is not universally suitable. It is restricted to scenarios where communication among agents is limited. In contrast, real-world situations often involve agents with robust communication capabilities, enabling them to freely share observations. This sharing of observations enhances agents' awareness to the environment and can lead to better cooperation, while decentralized execution neglects this. Second, there exist situations where the agents' actions exhibit interdependence. Agents in a system can sometimes be divided into dominant agents and assistant agents~\cite{hete_graph,first_move,master_slave,leader_following}. The actions of the dominant agents carry greater importance, granting them priority in decision-making. Subsequently, the assistant agents make decisions based on the dominant agents' actions, playing a supportive role. Such a cooperative pattern is also common in human society, where individual actions are not independent but rather correlated, challenging the independence assumption.

\begin{figure}[htbp]
\centering
\includegraphics[width=0.7\linewidth]{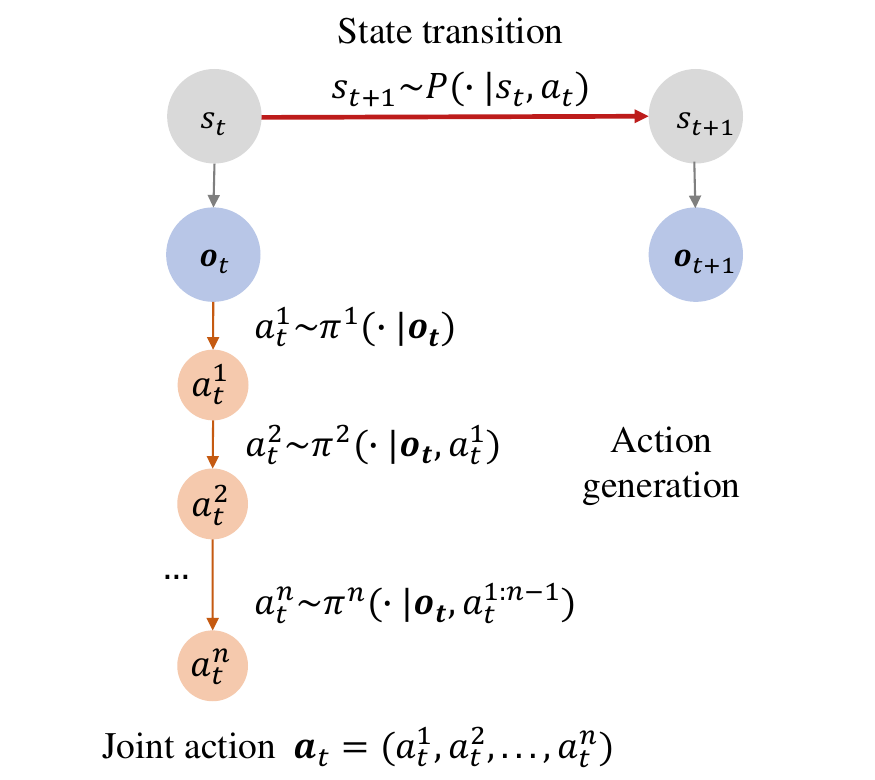}
\caption{Action generation process.}
\label{fig2}
\end{figure}

\begin{figure*}
\centering
\includegraphics[width=0.8\linewidth]{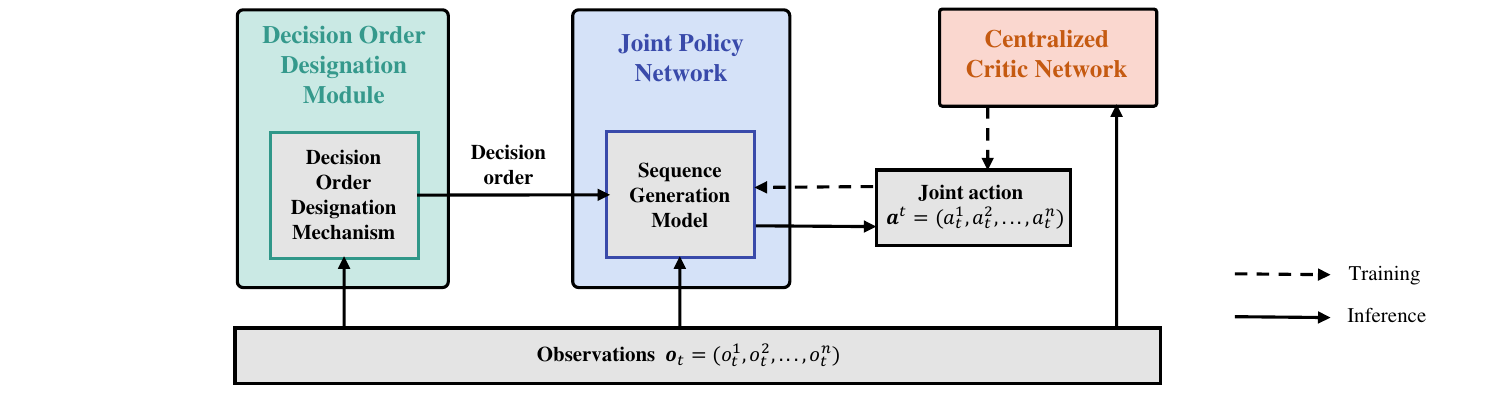}
\caption{Illustration of the general framework of solving MARL using sequence generation model.}
\label{fig3}
\end{figure*}

Therefore, we propose an alternative decomposition method for the joint policy that does not rely on the assumption of independence among agents' actions. Formally, we decompose the joint policy into conditional probabilities:
\begin{equation}
\begin{aligned}
    \boldsymbol{\pi}\left(\boldsymbol{a}_t|\boldsymbol{o}_t\right) &= \boldsymbol{\pi}\left(a^1_t, a^2_t,\dots,a^n_t|\boldsymbol{o}_t\right) \\
    &=\prod_{i=1}^n\pi^i\left(a^i_t|\boldsymbol{o}_t,a^{1:i-1}_t\right),\label{eq3}
\end{aligned}
\end{equation}
where $\pi^i\left(a^i_t|\boldsymbol{o}_t,a^{1:i-1}_t\right)$, called the conditional individual policy, is the conditional probability of the $i^{th}$ agent's action given the joint observation and the preceding agents' actions. In this way, the decision-making process of MAS is explicitly transformed into a sequence generation task: given the joint observation $\boldsymbol{o}_t$ at each time step, the actions of agents are generated sequentially. This generation process are illustrated in Figure \ref{fig2}. 

Any sequence generation model has the potential to tackle this task. Besides, the order of generation is crucial in this process. Consequently, we propose a generalized framework composed of a Decision Order Designation Module, a Joint Policy Network and a Centralized Critic Network, to offer an effective pipeline within the sequence generation scheme. The proposed framework is illustrated in Figure \ref{fig3}. In the framework, the decision order designation module built with specific mechanism is used to designate the order of generation based on the agents' observations. The joint policy network is then used to generate agents' actions in the specified order, and is trained based on the value function approximated by the centralized critic network. 

This framework covers all the components necessary for generating agents' actions in a sequential manner, and offers the benefit of simplifying MARL into single-agent RL. The introduction of powerful sequence generation model also contributes to handling the exponential growth of the joint action space. We summarize our proposed framework in \ref{alg:Framework}.

\begin{algorithm}[htbp]
\caption{A General Framework of Solving MARL Using Sequence Generation Model}
\label{alg:Framework}
\textbf{Input}: Number of agents $n$.\\
\textbf{Initialize}: A centralized critic, a sequence generation model, a decision order designation mechanism, and a replay buffer.
\begin{algorithmic}[1]
\REPEAT
\STATE At each time step, collect agents' observations and then designate the action generation order $\left(i_1,\dots,i_n\right)$ based on the decision order designation mechanism.
\STATE  Interact with the environment with the joint policy network, which takes all agents' observation as input and generates $n$ actions sequentially in the specified order. Collect interaction data and input the data into replay buffer.
\STATE Train the centralized critic using the sampled data from the replay buffer to approximate the value function. 
\STATE Train the joint policy network using the sampled date and the value function approximated by centralized critic.
\UNTIL{The desired performance is achieved.}
\end{algorithmic}
\textbf{Output}: A trained joint policy network.
\end{algorithm}

\subsection{Transformer-Based Joint Policy Network}
Having transformed the decision-making process into a sequence generation task, we can take advantage of any state-of-the-art sequence models. The Transformer model~\cite{transformer}, which was originally designed for machine translation tasks, has exhibited strong sequence modeling abilities. Hence, we adopt the Transformer architecture introduced in MAT to construct our joint policy network.

\begin{figure*}
\centering
\includegraphics[width=0.8\linewidth]{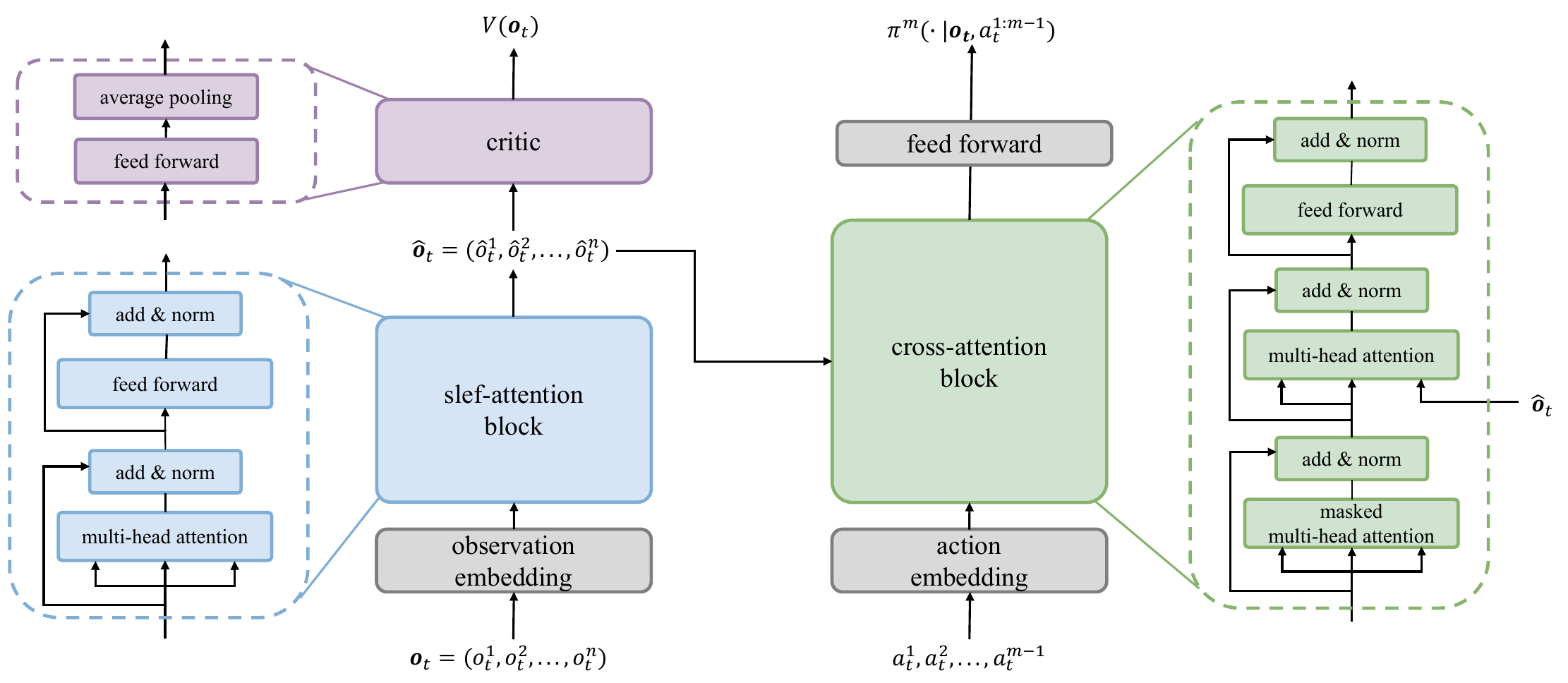}
\caption{Architecture of the Transformer-based policy network.}
\label{fig4}
\end{figure*}

As illustrated in Figure \ref{fig4} , our Transformer-based joint policy network consists of an \textit{encoder}, a \textit{decoder}, and a centralized critic. The \textit{encoder}, whose parameters are denoted by $\phi$, plays a vital role of learning an effective representation of the original observations $\boldsymbol{\hat{o}}_t=\left(\hat{o}^1_t,\dots,\hat{o}^n_t\right)$. This is achieved through a self-attention block consisting of a self-attention mechanism, Multi-Layer Perceptrons (MLPs), and residual connections. Such computational block takes all the agents' original observations $\boldsymbol{o}_t=\left(o^1_t,\dots,o^i_t\right)$ as input, integrates task-related information and outputs the encoded observations $\boldsymbol{\hat{o}}_t=\left(\hat{o}^1_t,\dots,\hat{o}^n_t\right)$. Then those coded observations are passed through the centralized critic, whose parameters are denoted by $\psi$, to calculate the joint observation value $V_\psi\left(\boldsymbol{\hat{o}}_t\right)$, which is then used to calculate the joint PPO loss. 

The \textit{decoder}, whose parameters are denoted by $\theta$, acts as a sequence generator that generates the agent's action $\boldsymbol{a}_t=\left(a^1_t,\dots,a^i_t\right)$ in an auto-regression manner. Specifically, this process begins with an input of a initial token as well as the encoded observation $\boldsymbol{\hat{o}_t}$, and output of the first agent's conditional individual policy $\pi^1_\theta\left(a^1_t|\boldsymbol{\hat{o}}_t\right)$, which is actually the probability distribution over possible actions. The first agent's action $a^1_t$ is sampled from this distribution, encoded as a one-hot vector, and then fed back into the \textit{decoder} as the second token. Subsequently, the second agent's action $a^2_t$ is sampled according to the output conditional individual policy $\pi^2_\theta\left(a^2_t|\boldsymbol{\hat{o}}_t,a^1_t\right)$. This process continues until all the agents' actions are generated, together forming the joint action. The \textit{decoder} block consists of a masked self-attention mechanism, a masked cross-attention mechanism, MLPs and some residual connections. The masked cross-attention mechanism is used to integrate the encoded observation, where ‘‘masked’’ indicates that the attention matrix is an upper triangular matrix ensuring that each agent's action $a^i_t$ depends only on its preceding generated actions $a^{j,j<i}_t$.

\subsection{Joint PPO Loss}
For network training, we use PPO algorithm to directly optimize the joint policy. To achieve this, an accurate approximation of the joint observation value function $V\left(\boldsymbol{o}_t\right)$ is necessary, so we build the loss functions for the critic and the policy separately. For the critic's loss, We first use the joint observation value function $V_\psi\left(\boldsymbol{\hat{o}}_{t}\right)$, which is approximated by the centralized critic, to estimate the joint advantage function, following the \textit{Generalized Advantage Estimation} (GAE)~\cite{gae} as:
\begin{equation}
    A\left(\boldsymbol{o}_t, \boldsymbol{a}_t\right)=\sum^{h}_{l=0}\left(\gamma\lambda\right)^l\delta_{t+l},
\end{equation}
where $\delta_t=r_t+\gamma V_\psi\left(\boldsymbol{\hat{o}}_{t+1}\right)-V_{\psi}\left(\boldsymbol{\hat{o}}_t\right)$ is the TD error at time step $t$ and $h$ is GAE steps. Similar to IPPO~\cite{ippo}, we formulate the critic's loss as the error between the predicted joint observation value $V_{\psi}\left(\boldsymbol{\hat{o}}_t\right)$ and its estimated value based on the real collected rewards:
\begin{equation}
\begin{aligned}
      \mathcal{L}_{critic}=  \frac{1}{T}\sum^{T-1}_{t=0}\min\left[\left(V_\psi\left(\boldsymbol{\hat{o}}_t\right)-\hat{V}_t\right)^2,\Big(V_{\psi_{old}}\left(\boldsymbol{\hat{o}}_t\right)+\right. \\
      \left.clip\left(V_\psi\left(\boldsymbol{\hat{o}}_t\right)-V_{\psi_{old}}\left(\boldsymbol{\hat{o}}_t\right),-\epsilon,+\epsilon\right)-\hat{V}_t\Big)^2\right],\label{eq5}
\end{aligned}     
\end{equation}
where $\hat{V}_t=A_t+V_\psi(\boldsymbol{\hat{o}}_t)$ and $\psi_{\theta_{old}}$ are the old parameters before the update. Eq.~\eqref{eq5} restricts the update of the centralized value function to within a trust region, preventing from overfitting to previous data as the data distribution continually changes with the evolving policy. Since the critic takes the encoded observation $\boldsymbol{\hat{o}}_t$ as input, this loss also contributes to the training of the \textit{encoder} to learn an expressive representation of the observations. 

As for policy training, we employ PPO on the generated joint policy. The joint policy $\boldsymbol{\pi}_\theta\left(\boldsymbol{a}_t|\boldsymbol{\hat{o}}_t\right)$ is computed following Eq.~\eqref{eq3} by multiplying the generated conditional local policies. Formally, the joint PPO loss is constructed as follows:
\begin{subequations}
\begin{align}
    \mathcal{L}_{policy}=-\frac{1}{T}\sum^{T-1}_{t=0}\min\Big(\alpha_tA_t,clip\left(\alpha_t,1\pm\epsilon\right)A_t\Big),\label{eq6a}
\end{align}
where
\begin{align}
    \alpha_t &= \frac{\boldsymbol{\pi}_\theta\left(\boldsymbol{a}_t|\boldsymbol{\hat{o}}_t\right)}{\boldsymbol{\pi}_{\theta_{old}}\left(\boldsymbol{a}_t|\boldsymbol{\hat{o}}_t\right)}\label{eq6b} \\
    &=\frac{\prod_{i=1}^n\pi^i_\theta\left(a^i_t|\boldsymbol{\hat{o}}_t,a^{1:i-1}_t\right)}{\prod_{i=1}^n\pi^i_{\theta_{old}}\left(a^i_t|\boldsymbol{\hat{o}}_t,a^{1:i-1}_t\right)} .\nonumber
\end{align}
\end{subequations}
Eq.~\eqref{eq6a} presents a direct use of PPO on the joint policy $\boldsymbol{\pi}_\theta\left(\boldsymbol{a}_t|\boldsymbol{\hat{o}}_t\right)$, whose update is restricted to within the trust region. In this way, JointPPO operates on all variables at the joint level, making it a fully centralized method. The theoretical advantage of PPO also facilitates the monotonic improvement of the joint policy. 

Having constructed the loss function for both the critic and the policy network, the overall learning loss can be computed by:
\begin{equation}
    \mathcal{L}(\theta,\phi, \psi) = \mathcal{L}_{critic}+\lambda_{1}\mathcal{L}_{policy}+\lambda_{2}\sum^{n}_{i=1}\mathcal{H}(\pi^i_\theta),\label{eq7}
\end{equation}
where $\mathcal{H}(\pi^i_\theta)$ denotes the entropy of the conditional individual policy $\pi^i_\theta$, serving to prevent from early convergence to suboptimal solutions, and $\lambda_{1}$,$\lambda_{2}$ are the weight parameters. 

\subsection{Decision Order Designation Mechanism}

In our proposed framework, a decision order designation mechanism is necessary to designate the order of action generation at each time step. In this paper, we present and study two kinds of mechanisms: the prior knowledge based mechanism and the graph generative model based mechanism.

Our first choice is to implement the prior knowledge based mechanism. In this approach, we manually designate the generation order of actions based on our prior knowledge to the task. Specifically, the types of agents are taken as the primary criterion for determining their decision order. This approach offers the advantage of simplicity and straightforwardness requiring no additional calculations. However, it lacks flexibility and may overlook potential dependencies among agents.
 
\begin{figure}[htbp]
\centering
\includegraphics[width=1\linewidth]{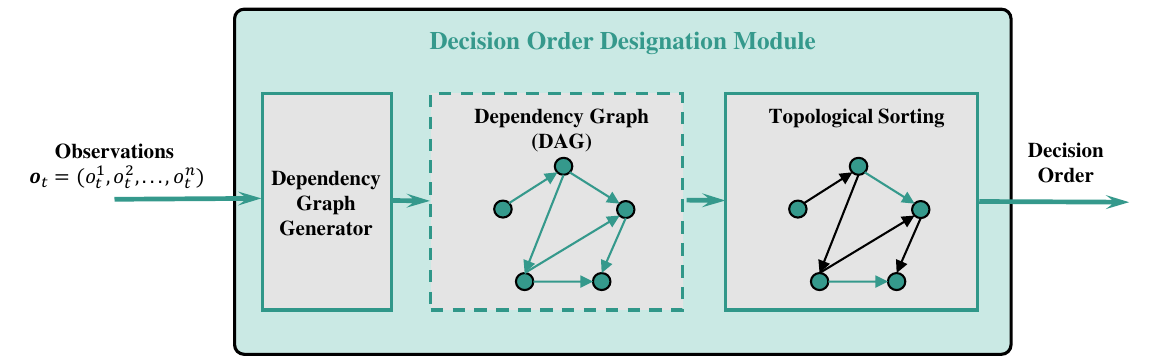}
\caption{Illustration of the graph generative model based mechanism.}
\label{fig5}
\end{figure}

\begin{table*}[t]
\caption{Performance evaluations of win rate and standard deviation on the SMAC testbed.}
\label{evaluation results}
\centering
\begin{tabular}{ccc|cccc|c}
\toprule
Task&Type&Difficulty&JointPPO&MAT&MAPPO&HAPPO&Steps \\
\midrule
5m\_vs\_6m&Homogeneous&Hard&\textbf{89.06} (0.03) & 72.81 (0.17) & 85.62 (0.05) & 61.56 (0.07) & 1e7 \\
8m\_vs\_9m&Homogeneous&Hard&\textbf{98.44} (0.01)&97.81 (0.02)&97.50 (0.01)& 65.31 (0.03)&5e6 \\
10m\_vs\_11m&Homogeneous&Hard&\textbf{99.69} (0.00)&98.12 (0.02)&93.75 (0.03)&75.31 (0.09)&5e6 \\
27m\_vs\_30m&Homogeneous&Super Hard&\textbf{100.00} (0.00)&95.63 (0.04)&91.25 (0.04)& 0.00 (0.00)&1e7 \\
6h\_vs\_8z&Homogeneous&Super Hard&92.5 (0.04)&\textbf{97.81} (0.01)&77.50 (0.17)&0.08 (0.05)&1e7 \\
MMM&Heterogeneous&Easy&\textbf{99.69} (0.01)&96.88 (0.00)&97.50 (0.02)&0.00 (0.00)&2e6 \\
3s5z&Heterogeneous&Hard&\textbf{96.88} (0.01)&92.19 (0.05)&96.25 (0.02)&31.56 (0.18)&3e6 \\
MMM2&Heterogeneous&Super Hard&\textbf{97.19} (0.02)&88.44 (0.08)& 89.06 (0.07)& 0.01 (0.01)&1e7 \\
3s5z\_vs\_3s6z&Heterogeneous&Super Hard&91.56 (0.03)&\textbf{95.63} (0.03)&62.81 (0.04)& 88.12 (0.07)&2e7 \\
\bottomrule
\end{tabular}
\end{table*}

\begin{figure*}
\centering
\subfigure[Learning curves of win rate.]{\label{fig6a}
\includegraphics[width=1\linewidth]{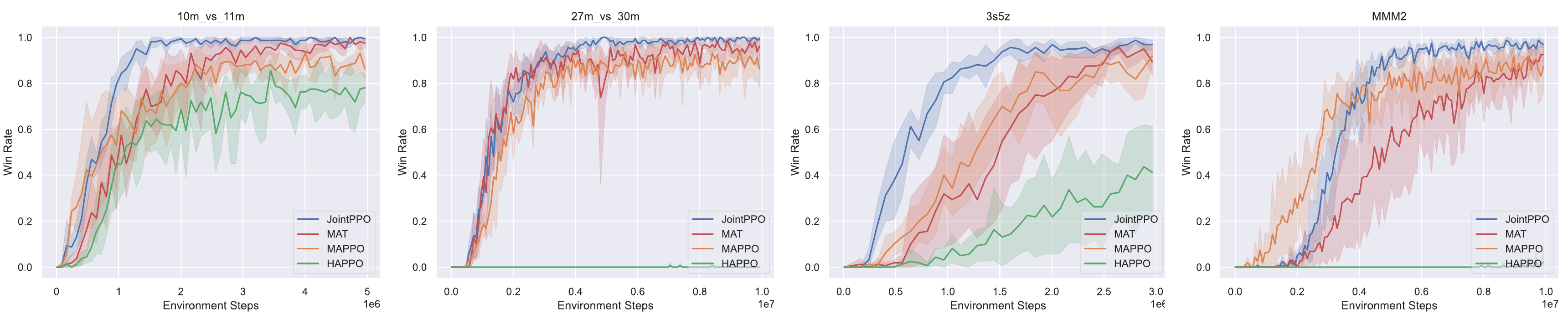}}\\
\subfigure[Learning curves of the number of killed allies.]{\label{fig6b}
\includegraphics[width=1\linewidth]{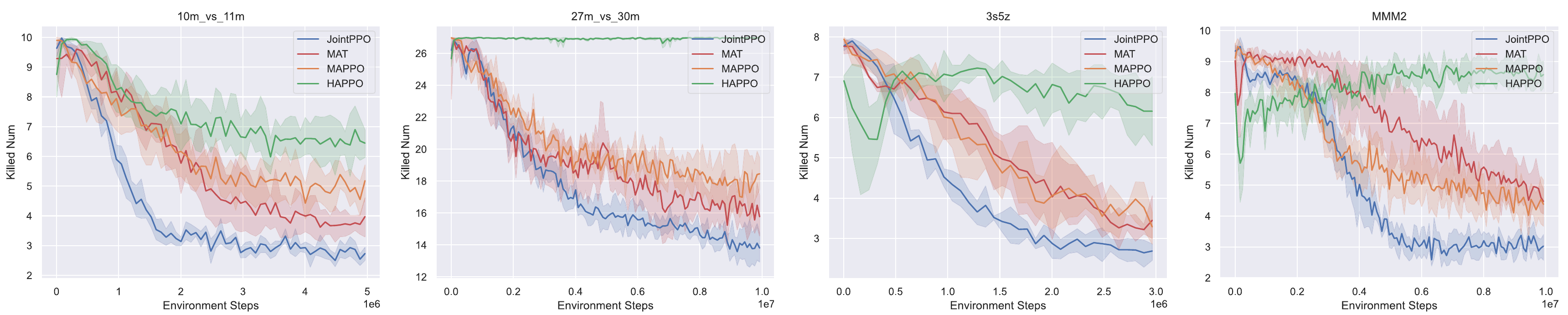}}
\caption{Comparison of JointPPO against baselines on four SMAC maps.}
\label{fig6}
\end{figure*}

We also implement a graph generative model based mechanism introduced in the work of~\cite{gcs}. In this approach, the multi-agent system is abstracted as a directed acyclic graph (DAG), where vertices represent agents and edges represent dependency among agents. A dependency graph generator based on the graph generative model is constructed and trained to generate the adjacency matrix of the dependency graph at each time step. By performing topological sorting on the generated dependency graph, the decision order of agents can be obtained. This process is illustrated in Figure \ref{fig5}.

The graph generative model based mechanism offers the advantage of flexibility by adjusting the decision order based on agents' real-time observation. However, the graph generator and topological sorting involve a significant amount of calculation, resulting in huge computational and time costs. Therefore, in the following experiments, we adopt the prior knowledge based mechanism as our default choice to validate the effectiveness of the joint PPO loss. Subsequently, we analyse and discuss the effects of different mechanisms in the ablation studies.


\section{Experiments}
In this section, we evaluate JointPPO across various tasks in the StarCraft Multi-Agent Challenge (SMAC) testbed. 

\subsection{SMAC Testbed}
SMAC (StarCraft Multi-Agent Challenge)~\cite{smac} is a testbed for MARL that offers a diverse set of StarCraft II unit micromanagement tasks of varying difficulties. In these tasks, a collaborative team of algorithm controlled units need to defeat an enemy team of units controlled by the built-in AI. The units in SMAC are also diverse. In homogeneous tasks, the units comprising a team are of the same type, whereas heterogeneous tasks mean the opposite. Successful strategies often require precise coordination among the units, executing tactics such as focused attack or kiting to gain positional advantages. 

For our experiments, we use game version 4.6, and all the evaluation results are averaged over 5 random seeds. For each random seed, following the evaluation metric proposed in~\cite{rode}, we compute the \textbf{win rates} over 32 evaluation games after each training iteration and take the median of the final ten evaluation win rates as the performance for each seed. However, evaluating algorithms solely based on win rates is not enough, as there exit situations that two algorithm with same win rates may differ in terms of the cost paid for the victory, such as the number of killed allies. Therefore, we further record the number of \textbf{killed allies} in the evaluation game as an additional performance metric. More details are presented in supplementary materials.

\subsection{JointPPO's Performance}

We present the performance of JointPPO on several representative tasks, covering both homogeneous and homogeneous settings. We compare JointPPO with PPO-based algorithms MAPPO, HAPPO, and SOTA algorithm MAT. We use the same hyperparameters of these baseline algorithms from their original papers to ensure their best performance and fair comparisons. The evaluation results and learning curves of win rates are presented in Table \ref{evaluation results} and Figure \ref{fig6a}. JointPPO exhibits competitive performance with baseline algorithms in terms of final win rates and sample efficiency. Remarkably, it achieves nearly 100\% test win rates across all test maps, including the super hard heterogeneous task MMM2, which is not easily solved by existing methods. Additionally, we surprisingly observe that JointPPO demonstrates a lower cost of killed allied for victory in most tasks (Figure \ref{fig6b}). This indicates that the joint policy converges to a better equilibrium, and we attribute this to JointPPO's direct optimization of the joint policy. 

\subsection{Ablation Studies}
\begin{figure*}
\centering
\subfigure[]{\label{fig7a}
\includegraphics[width=0.24\linewidth]{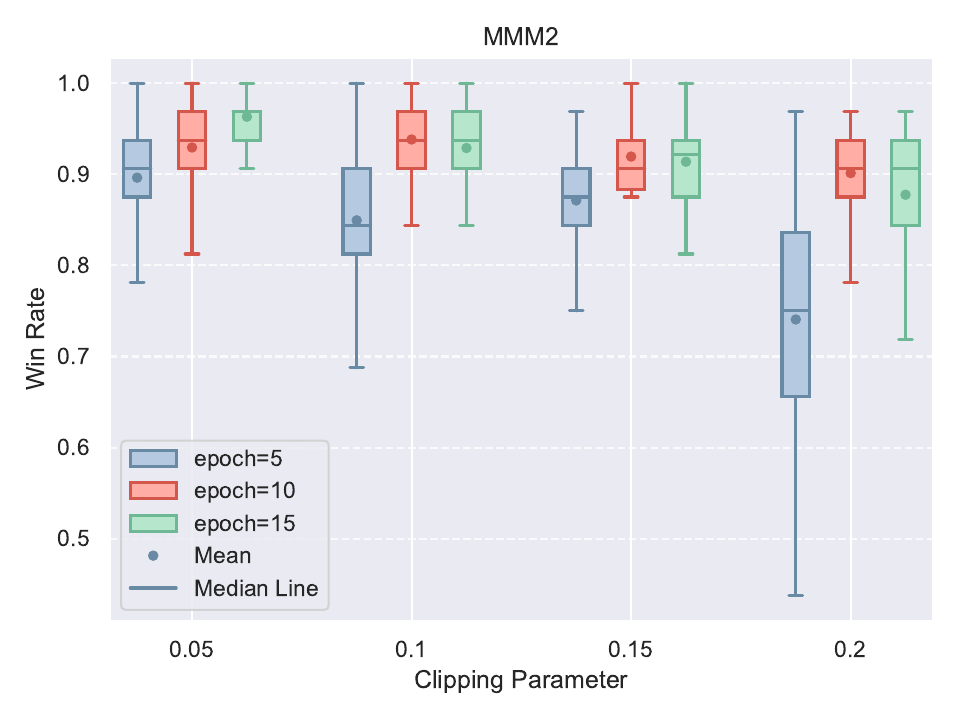}}\hspace{-3mm}
\subfigure[]{\label{fig7b}
\includegraphics[width=0.25\linewidth]{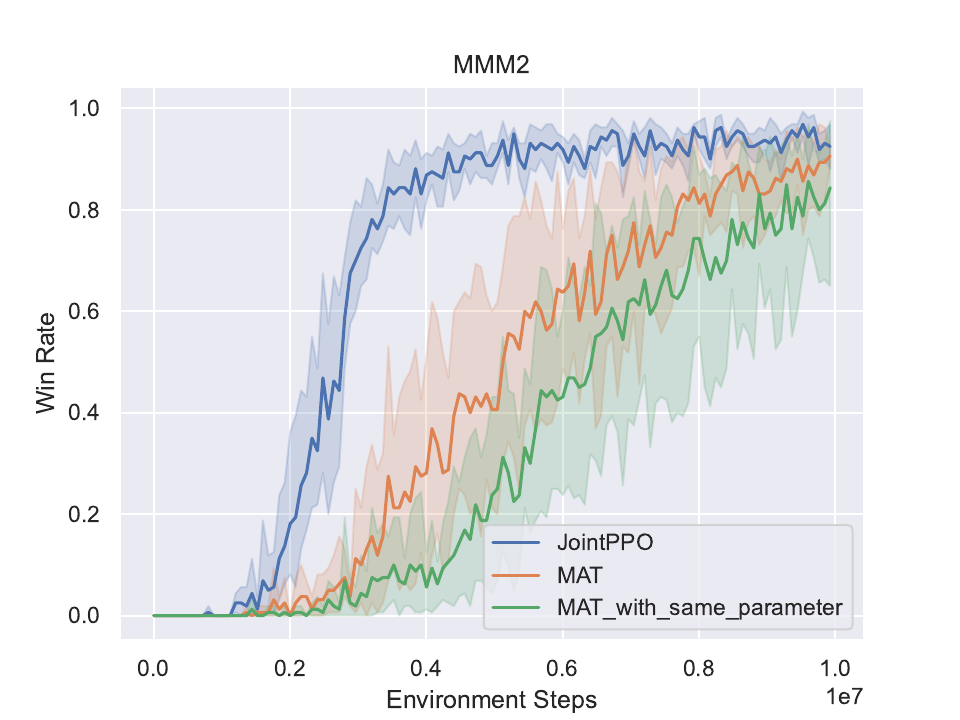}}\hspace{-5mm}
\subfigure[]{\label{fig7c}
\includegraphics[width=0.25\linewidth]{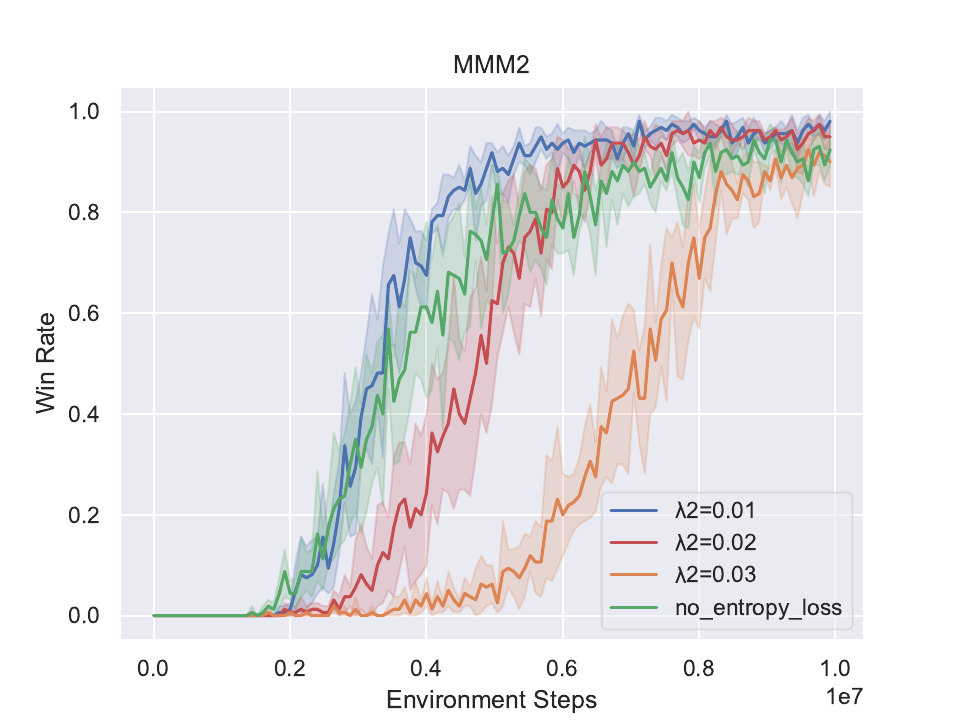}}\hspace{-5mm}
\subfigure[]{\label{fig7d}
\includegraphics[width=0.25\linewidth]{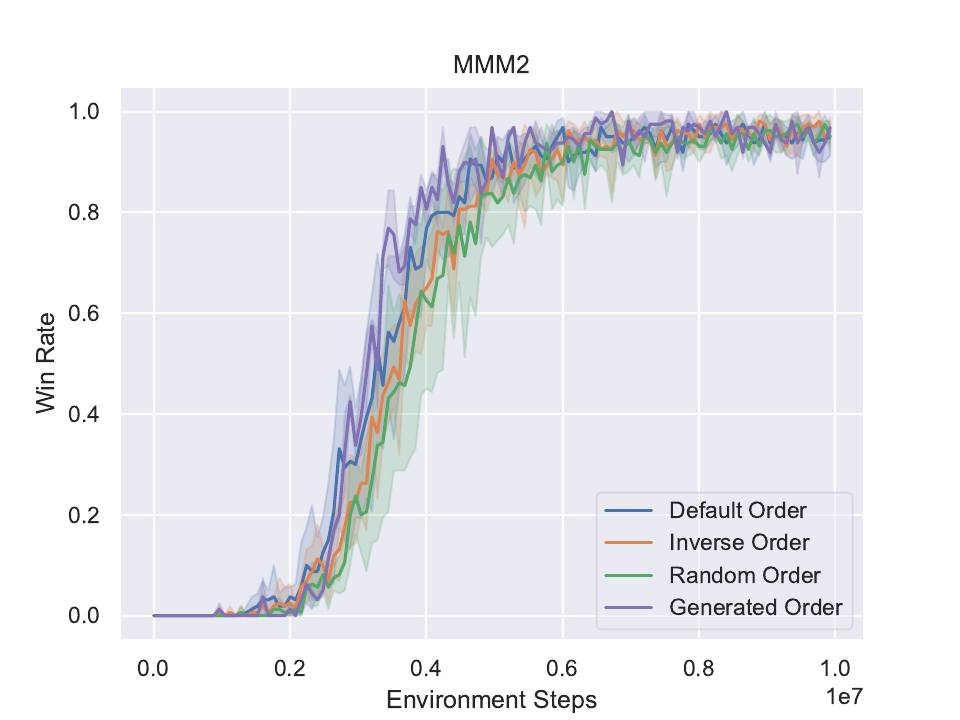}}
\caption{Results of ablation experiments. (a): Effect of training epochs and clipping parameter.(b): Comparison between JointPPO, original MAT and MAT with same hyperparameter. (c): Comparison of JointPPO with different weighted entropy loss. (d): Comparison of JointPPO with different decision order designation mechanism.}
\label{fig7}
\end{figure*}
We conduct ablation experiments and analyses on factors that's important in JointPPO's implementation including: PPO training epochs and clipping parameter, entropy loss, and decision order designation mechanism. Each factor is studies through a series of experiments on the super hard heterogeneous tasks MMM2.

\subsubsection{\textbf{PPO Training Epochs and Clipping Parameter}}
These two parameters are among the most influential hyperparameters affecting JointPPO's training. We conduct experiments with different combination of these two parameters. The results are presented in Figure \ref{fig7a}. Generally, the final win rates are positively correlated with training epochs, while negatively correlated with the clipping parameter. However, as the training epoch further increases, the training process will crush, which was observed during experiments. This trend validates the explanation in \cite{mappo} that greater training epochs bring higher sample efficiency, but may hurt the optimality of convergence due to an overfitting on old data. A larger clipping parameter may cause instability, as seen in the experiment with {epoch=5,clipping parameter=0.2}. 

We further conduct experiments using MAT with the same training epochs and clipping parameter as JointPPO. We compare its performance with JointPPO and original MAT in Figure \ref{fig7b}. The results show that MAT with the same parameters perform worse than the its original settings, thereby eliminating the impact of parameter tuning on the JointPPO's improved performance.

\subsubsection{\textbf{Entropy Loss}}
The entropy loss is a crucial component contributing to the learning loss. Its strength is determined by the weight parameter $\lambda_2$ (Eq.~\ref{eq7}). A well tuned $\lambda_2$ is supposed to strike a balance between exploration and exploitation. Here, we investigate how entropy loss with different weights affects the training process. Figure \ref{fig7c} shows that, a smaller weight can result in faster convergence while a larger one does the opposite. However, when it's set too small, such as an extreme case, zero, it brings a lower final win rate which indicates a suboptimal convergence. Therefore, in all of our experiment, the weight $\lambda_2$ is set as an compromise as 0.1.

\subsubsection{\textbf{Decision Order Designation Mechanism}}
The decision order designation mechanism plays a vital role in designating the generation order of agents' actions. Here we investigate how this order affects JointPPO's performance. We conduct experiments with two mentioned mechanism: the prior knowledge based mechanism and the graph generated based mechanism. For the former, we further compare difference orders \textemdash specifically, the inverse order and random order \textemdash to the default one. The experiments results presented in Figure \ref{fig7d} shows that JointPPO achieves high win rates across all generation orders, with the graph generative model based order showing slightly faster convergence than others. The results suggest that the graph generative model based mechanism exhibits higher sample efficiency due to its additional use of data to learn a dependency graph. Nevertheless, these four settings show minimal differences in final performance, demonstrating JointPPO's robustness to generation order.


\section{Conclusion}
In this paper, we decompose the joint policy of the multi-agent system into conditional probabilities and introduce a framework that solves MARL using the sequence generation model. By leveraging the Transformer model, the proposed CTCE algorithm, JointPPO, effectively handles high-dimensional joint action spaces and demonstrates competitive performance on SMAC. Furthermore, it demonstrates the feasibility of addressing MARL from a fully centralized perspective and reveals promising possibilities for integration with promising advancements in single-agent RL domain, which we leave for future study.

\bibliographystyle{named}
\bibliography{ijcai24}

\begin{thebibliography}{}

\bibitem[\protect\citeauthoryear{Bertsekas}{2021}]{rpisa}
Dimitri Bertsekas.
\newblock Multiagent reinforcement learning: Rollout and policy iteration.
\newblock {\em IEEE/CAA Journal of Automatica Sinica}, 8(2):249--272, 2021.

\bibitem[\protect\citeauthoryear{Chen \bgroup \em et al.\egroup }{2022}]{csrl}
Yiqun Chen, Wei Yang, Tianle Zhang, Shiguang Wu, and Hongxing Chang.
\newblock Commander-soldiers reinforcement learning for cooperative multi-agent systems.
\newblock In {\em 2022 International Joint Conference on Neural Networks (IJCNN)}, pages 1--7. IEEE, 2022.

\bibitem[\protect\citeauthoryear{de Witt \bgroup \em et al.\egroup }{2020}]{ippo}
Christian~Schroeder de~Witt, Tarun Gupta, Denys Makoviichuk, Viktor Makoviychuk, Philip~HS Torr, Mingfei Sun, and Shimon Whiteson.
\newblock Is independent learning all you need in the starcraft multi-agent challenge?
\newblock {\em arXiv preprint arXiv:2011.09533}, 2020.

\bibitem[\protect\citeauthoryear{Dorri \bgroup \em et al.\egroup }{2018}]{mas}
Ali Dorri, Salil~S Kanhere, and Raja Jurdak.
\newblock Multi-agent systems: A survey.
\newblock {\em Ieee Access}, 6:28573--28593, 2018.

\bibitem[\protect\citeauthoryear{Du \bgroup \em et al.\egroup }{2022}]{hete_graph}
Wei Du, Shifei Ding, Chenglong Zhang, and Zhongzhi Shi.
\newblock Multiagent reinforcement learning with heterogeneous graph attention network.
\newblock {\em IEEE Transactions on Neural Networks and Learning Systems}, 2022.

\bibitem[\protect\citeauthoryear{Foerster \bgroup \em et al.\egroup }{2018}]{coma}
Jakob Foerster, Gregory Farquhar, Triantafyllos Afouras, Nantas Nardelli, and Shimon Whiteson.
\newblock Counterfactual multi-agent policy gradients.
\newblock In {\em Proceedings of the AAAI conference on artificial intelligence}, volume~32, 2018.

\bibitem[\protect\citeauthoryear{Han \bgroup \em et al.\egroup }{2020}]{robot_coordination}
Ruihua Han, Shengduo Chen, and Qi~Hao.
\newblock Cooperative multi-robot navigation in dynamic environment with deep reinforcement learning.
\newblock In {\em 2020 IEEE International Conference on Robotics and Automation (ICRA)}, pages 448--454. IEEE, 2020.

\bibitem[\protect\citeauthoryear{Kaelbling \bgroup \em et al.\egroup }{1998}]{pomdp}
Leslie~Pack Kaelbling, Michael~L Littman, and Anthony~R Cassandra.
\newblock Planning and acting in partially observable stochastic domains.
\newblock {\em Artificial intelligence}, 101(1-2):99--134, 1998.

\bibitem[\protect\citeauthoryear{Kakade and Langford}{2002}]{theoretical_guarantee}
Sham Kakade and John Langford.
\newblock Approximately optimal approximate reinforcement learning.
\newblock In {\em Proceedings of the Nineteenth International Conference on Machine Learning}, pages 267--274, 2002.

\bibitem[\protect\citeauthoryear{Kraemer and Banerjee}{2016}]{ctde2}
Landon Kraemer and Bikramjit Banerjee.
\newblock Multi-agent reinforcement learning as a rehearsal for decentralized planning.
\newblock {\em Neurocomputing}, 190:82--94, 2016.

\bibitem[\protect\citeauthoryear{Kuba \bgroup \em et al.\egroup }{2021}]{happo}
Jakub~Grudzien Kuba, Ruiqing Chen, Muning Wen, Ying Wen, Fanglei Sun, Jun Wang, and Yaodong Yang.
\newblock Trust region policy optimisation in multi-agent reinforcement learning.
\newblock {\em arXiv preprint arXiv:2109.11251}, 2021.

\bibitem[\protect\citeauthoryear{Kuba \bgroup \em et al.\egroup }{2022}]{mirror_learning}
Jakub~Grudzien Kuba, Xidong Feng, Shiyao Ding, Hao Dong, Jun Wang, and Yaodong Yang.
\newblock Heterogeneous-agent mirror learning: A continuum of solutions to cooperative marl.
\newblock {\em arXiv preprint arXiv:2208.01682}, 2022.

\bibitem[\protect\citeauthoryear{Lee \bgroup \em et al.\egroup }{2007}]{finance}
Jae~Won Lee, Jonghun Park, O~Jangmin, Jongwoo Lee, and Euyseok Hong.
\newblock A multiagent approach to $ q $-learning for daily stock trading.
\newblock {\em IEEE Transactions on Systems, Man, and Cybernetics-Part A: Systems and Humans}, 37(6):864--877, 2007.

\bibitem[\protect\citeauthoryear{Li \bgroup \em et al.\egroup }{2021}]{mamt}
Wenhao Li, Xiangfeng Wang, Bo~Jin, Junjie Sheng, and Hongyuan Zha.
\newblock Dealing with non-stationarity in marl via trust-region decomposition.
\newblock {\em arXiv preprint arXiv:2102.10616}, 2021.

\bibitem[\protect\citeauthoryear{Li \bgroup \em et al.\egroup }{2023}]{ace}
Chuming Li, Jie Liu, Yinmin Zhang, Yuhong Wei, Yazhe Niu, Yaodong Yang, Yu~Liu, and Wanli Ouyang.
\newblock Ace: Cooperative multi-agent q-learning with bidirectional action-dependency.
\newblock In {\em Proceedings of the AAAI conference on artificial intelligence}, volume~37, pages 8536--8544, 2023.

\bibitem[\protect\citeauthoryear{Liu \bgroup \em et al.\egroup }{2021}]{copa}
Bo~Liu, Qiang Liu, Peter Stone, Animesh Garg, Yuke Zhu, and Anima Anandkumar.
\newblock Coach-player multi-agent reinforcement learning for dynamic team composition.
\newblock In {\em International Conference on Machine Learning}, pages 6860--6870. PMLR, 2021.

\bibitem[\protect\citeauthoryear{Lowe \bgroup \em et al.\egroup }{2017}]{maddpg}
Ryan Lowe, Yi~I Wu, Aviv Tamar, Jean Harb, OpenAI Pieter~Abbeel, and Igor Mordatch.
\newblock Multi-agent actor-critic for mixed cooperative-competitive environments.
\newblock {\em Advances in neural information processing systems}, 30, 2017.

\bibitem[\protect\citeauthoryear{Mahajan \bgroup \em et al.\egroup }{2019}]{maven}
Anuj Mahajan, Tabish Rashid, Mikayel Samvelyan, and Shimon Whiteson.
\newblock Maven: Multi-agent variational exploration.
\newblock {\em Advances in neural information processing systems}, 32, 2019.

\bibitem[\protect\citeauthoryear{Nguyen \bgroup \em et al.\egroup }{2020}]{review_of_challenge}
Thanh~Thi Nguyen, Ngoc~Duy Nguyen, and Saeid Nahavandi.
\newblock Deep reinforcement learning for multiagent systems: A review of challenges, solutions, and applications.
\newblock {\em IEEE transactions on cybernetics}, 50(9):3826--3839, 2020.

\bibitem[\protect\citeauthoryear{Oliehoek \bgroup \em et al.\egroup }{2008}]{ctde1}
Frans~A Oliehoek, Matthijs~TJ Spaan, and Nikos Vlassis.
\newblock Optimal and approximate q-value functions for decentralized pomdps.
\newblock {\em Journal of Artificial Intelligence Research}, 32:289--353, 2008.

\bibitem[\protect\citeauthoryear{Rashid \bgroup \em et al.\egroup }{2020a}]{wqmix}
Tabish Rashid, Gregory Farquhar, Bei Peng, and Shimon Whiteson.
\newblock Weighted qmix: Expanding monotonic value function factorisation for deep multi-agent reinforcement learning.
\newblock {\em Advances in neural information processing systems}, 33:10199--10210, 2020.

\bibitem[\protect\citeauthoryear{Rashid \bgroup \em et al.\egroup }{2020b}]{qmix}
Tabish Rashid, Mikayel Samvelyan, Christian~Schroeder De~Witt, Gregory Farquhar, Jakob Foerster, and Shimon Whiteson.
\newblock Monotonic value function factorisation for deep multi-agent reinforcement learning.
\newblock {\em The Journal of Machine Learning Research}, 21(1):7234--7284, 2020.

\bibitem[\protect\citeauthoryear{Ruan \bgroup \em et al.\egroup }{2022a}]{gcs}
Jingqing Ruan, Yali Du, Xuantang Xiong, Dengpeng Xing, Xiyun Li, Linghui Meng, Haifeng Zhang, Jun Wang, and Bo~Xu.
\newblock Gcs: graph-based coordination strategy for multi-agent reinforcement learning.
\newblock {\em arXiv preprint arXiv:2201.06257}, 2022.

\bibitem[\protect\citeauthoryear{Ruan \bgroup \em et al.\egroup }{2022b}]{first_move}
Jingqing Ruan, Linghui Meng, Xuantang Xiong, Dengpeng Xing, and Bo~Xu.
\newblock Learning multi-agent action coordination via electing first-move agent.
\newblock In {\em Proceedings of the International Conference on Automated Planning and Scheduling}, volume~32, pages 624--628, 2022.

\bibitem[\protect\citeauthoryear{Samvelyan \bgroup \em et al.\egroup }{2019}]{smac}
Mikayel Samvelyan, Tabish Rashid, Christian~Schroeder De~Witt, Gregory Farquhar, Nantas Nardelli, Tim~GJ Rudner, Chia-Man Hung, Philip~HS Torr, Jakob Foerster, and Shimon Whiteson.
\newblock The starcraft multi-agent challenge.
\newblock {\em arXiv preprint arXiv:1902.04043}, 2019.

\bibitem[\protect\citeauthoryear{Schrittwieser \bgroup \em et al.\egroup }{2020}]{muzero}
Julian Schrittwieser, Ioannis Antonoglou, Thomas Hubert, Karen Simonyan, Laurent Sifre, Simon Schmitt, Arthur Guez, Edward Lockhart, Demis Hassabis, Thore Graepel, et~al.
\newblock Mastering atari, go, chess and shogi by planning with a learned model.
\newblock {\em Nature}, 588(7839):604--609, 2020.

\bibitem[\protect\citeauthoryear{Schulman \bgroup \em et al.\egroup }{2015a}]{trpo}
John Schulman, Sergey Levine, Pieter Abbeel, Michael Jordan, and Philipp Moritz.
\newblock Trust region policy optimization.
\newblock In {\em International conference on machine learning}, pages 1889--1897. PMLR, 2015.

\bibitem[\protect\citeauthoryear{Schulman \bgroup \em et al.\egroup }{2015b}]{gae}
John Schulman, Philipp Moritz, Sergey Levine, Michael Jordan, and Pieter Abbeel.
\newblock High-dimensional continuous control using generalized advantage estimation.
\newblock {\em arXiv preprint arXiv:1506.02438}, 2015.

\bibitem[\protect\citeauthoryear{Schulman \bgroup \em et al.\egroup }{2017}]{ppo}
John Schulman, Filip Wolski, Prafulla Dhariwal, Alec Radford, and Oleg Klimov.
\newblock Proximal policy optimization algorithms.
\newblock {\em arXiv preprint arXiv:1707.06347}, 2017.

\bibitem[\protect\citeauthoryear{Son \bgroup \em et al.\egroup }{2019}]{qtran}
Kyunghwan Son, Daewoo Kim, Wan~Ju Kang, David~Earl Hostallero, and Yung Yi.
\newblock Qtran: Learning to factorize with transformation for cooperative multi-agent reinforcement learning.
\newblock In {\em International conference on machine learning}, pages 5887--5896. PMLR, 2019.

\bibitem[\protect\citeauthoryear{Sunehag \bgroup \em et al.\egroup }{2017}]{vdn}
Peter Sunehag, Guy Lever, Audrunas Gruslys, Wojciech~Marian Czarnecki, Vinicius Zambaldi, Max Jaderberg, Marc Lanctot, Nicolas Sonnerat, Joel~Z Leibo, Karl Tuyls, et~al.
\newblock Value-decomposition networks for cooperative multi-agent learning.
\newblock {\em arXiv preprint arXiv:1706.05296}, 2017.

\bibitem[\protect\citeauthoryear{Tampuu \bgroup \em et al.\egroup }{2017}]{iql}
Ardi Tampuu, Tambet Matiisen, Dorian Kodelja, Ilya Kuzovkin, Kristjan Korjus, Juhan Aru, Jaan Aru, and Raul Vicente.
\newblock Multiagent cooperation and competition with deep reinforcement learning.
\newblock {\em PloS one}, 12(4):e0172395, 2017.

\bibitem[\protect\citeauthoryear{Vaswani \bgroup \em et al.\egroup }{2017}]{transformer}
Ashish Vaswani, Noam Shazeer, Niki Parmar, Jakob Uszkoreit, Llion Jones, Aidan~N Gomez, {\L}ukasz Kaiser, and Illia Polosukhin.
\newblock Attention is all you need.
\newblock {\em Advances in neural information processing systems}, 30, 2017.

\bibitem[\protect\citeauthoryear{Wang \bgroup \em et al.\egroup }{2020a}]{qplex}
Jianhao Wang, Zhizhou Ren, Terry Liu, Yang Yu, and Chongjie Zhang.
\newblock Qplex: Duplex dueling multi-agent q-learning.
\newblock {\em arXiv preprint arXiv:2008.01062}, 2020.

\bibitem[\protect\citeauthoryear{Wang \bgroup \em et al.\egroup }{2020b}]{rode}
Tonghan Wang, Tarun Gupta, Anuj Mahajan, Bei Peng, Shimon Whiteson, and Chongjie Zhang.
\newblock Rode: Learning roles to decompose multi-agent tasks.
\newblock {\em arXiv preprint arXiv:2010.01523}, 2020.

\bibitem[\protect\citeauthoryear{Wang \bgroup \em et al.\egroup }{2023a}]{master_slave}
Jiao Wang, Mingrui Yuan, Yun Li, and Zihui Zhao.
\newblock Hierarchical attention master--slave for heterogeneous multi-agent reinforcement learning.
\newblock {\em Neural Networks}, 162:359--368, 2023.

\bibitem[\protect\citeauthoryear{Wang \bgroup \em et al.\egroup }{2023b}]{a2po}
Xihuai Wang, Zheng Tian, Ziyu Wan, Ying Wen, Jun Wang, and Weinan Zhang.
\newblock Order matters: Agent-by-agent policy optimization.
\newblock {\em arXiv preprint arXiv:2302.06205}, 2023.

\bibitem[\protect\citeauthoryear{Wen \bgroup \em et al.\egroup }{2022}]{mat}
Muning Wen, Jakub Kuba, Runji Lin, Weinan Zhang, Ying Wen, Jun Wang, and Yaodong Yang.
\newblock Multi-agent reinforcement learning is a sequence modeling problem.
\newblock {\em Advances in Neural Information Processing Systems}, 35:16509--16521, 2022.

\bibitem[\protect\citeauthoryear{Wu \bgroup \em et al.\egroup }{2020}]{traffic_light}
Tong Wu, Pan Zhou, Kai Liu, Yali Yuan, Xiumin Wang, Huawei Huang, and Dapeng~Oliver Wu.
\newblock Multi-agent deep reinforcement learning for urban traffic light control in vehicular networks.
\newblock {\em IEEE Transactions on Vehicular Technology}, 69(8):8243--8256, 2020.

\bibitem[\protect\citeauthoryear{Wu \bgroup \em et al.\egroup }{2023}]{daydreamer}
Philipp Wu, Alejandro Escontrela, Danijar Hafner, Pieter Abbeel, and Ken Goldberg.
\newblock Daydreamer: World models for physical robot learning.
\newblock In {\em Conference on Robot Learning}, pages 2226--2240. PMLR, 2023.

\bibitem[\protect\citeauthoryear{Yu \bgroup \em et al.\egroup }{2022}]{mappo}
Chao Yu, Akash Velu, Eugene Vinitsky, Jiaxuan Gao, Yu~Wang, Alexandre Bayen, and Yi~Wu.
\newblock The surprising effectiveness of ppo in cooperative multi-agent games.
\newblock {\em Advances in Neural Information Processing Systems}, 35:24611--24624, 2022.

\bibitem[\protect\citeauthoryear{Zhang \bgroup \em et al.\egroup }{2022}]{leader_following}
Feiye Zhang, Qingyu Yang, and Dou An.
\newblock A leader-following paradigm based deep reinforcement learning method for multi-agent cooperation games.
\newblock {\em Neural Networks}, 156:1--12, 2022.

\bibitem[\protect\citeauthoryear{Zhang \bgroup \em et al.\egroup }{2023}]{stackelberg}
Bin Zhang, Lijuan Li, Zhiwei Xu, Dapeng Li, and Guoliang Fan.
\newblock Inducing stackelberg equilibrium through spatio-temporal sequential decision-making in multi-agent reinforcement learning.
\newblock {\em arXiv preprint arXiv:2304.10351}, 2023.

\end{thebibliography}


\newpage
\appendix
\section{Pseudocode of JointPPO}
\begin{algorithm}[htbp]
\caption{JointPPO}
\label{alg:jointppo}
\textbf{Input}: Number of agents $N$, batch size $B$, episodes $K$, steps per episode $T$.\\
\textbf{Initialize}: \textit{Encoder}'s parameters ${\phi}$, \textit{decoder}'s parameters $\theta$, replay buffer $\mathcal{B}$.
\begin{algorithmic}[1] 
\FOR{$k=0,1,\dots,K-1$}
\STATE Initialize the environment and start an episode.
\FOR{$t=0,1,\dots,T-1$}
\STATE Collect observations $\boldsymbol{o}_t=\left\{o^i_t\right\}^n_{i=1}$.
\STATE Input the collected observations to the decision order designation module, get the action generation order $\left(i_1,\dots,i_n\right)$.
\STATE Input the collected observations to the joint policy network, get the generated conditional local policies $\left\{\pi^i\left(a^{i}_t|\boldsymbol{o}_t,a^{1:i-1}_t\right)\right\}_{i=1:n}$ under specified order.
\STATE Sample agents' actions and execute the joint action $\boldsymbol{a}_t=\left(a^1_t,\dots,a^i_t\right)$ to the environment. Receive the team reward $r_t$ and stimulate the environment to the next state.
\STATE Insert tuple $\left(\boldsymbol{o}_t, V_\phi\left(\boldsymbol{o}_t\right), \left\{\pi^i\right\}, \boldsymbol{a}_t, r_t\right)$ in to $\mathcal{B}$.
\ENDFOR
\STATE Sample a random minibatch of $B$ from $\mathcal{B}$.
\STATE Calculate the loss according to Eq.~\eqref{eq7} and update network parameters ${\phi}$ and $\theta$ with gradient descent.
\STATE Update the graph generative model following \cite{gcs}.
\ENDFOR
\end{algorithmic}
\textbf{Output}: A trained Transformer-based joint policy network.
\end{algorithm}

\section{Hyper-parameter Settings for Experiments}
During experiments, the implementations of MAT, MAPPO and HAPPO are consistent with their official repositories. Here we list the hyper-parameter adopted in the implementation of JointPPO for different tasks in Table \ref{common hyper parameters} and Table \ref{different hyper parameters}, especially in terms of the ppo epochs, ppo clip, learning rate decay strategy, and the coefficient parameter $\lambda_1$ of the PPO loss, which balances the proportion of the entropy term in the overall learning loss. 

\begin{table}[htbp]
\caption{Common hyper-parameters used for JointPPO in the experiments.}
\label{common hyper parameters}
\centering
\begin{tabular}{c|c}
\toprule
hyper-parameters&value \\
\midrule
learning rate& 5e-4\\
batch size&3200 \\
discount factor&0.99\\
entropy coef& 0.01\\
hidden layer num&1\\
hidden layer dim&64\\
attention block num&1\\
optimizer&Adam \\
learning rate decay&True \\
use value normalization&True\\
\bottomrule
\end{tabular}
\end{table}

\begin{table}[htbp]
\caption{Different hyper-parameters used for JointPPO in the experiments.}
\label{different hyper parameters}
\centering
\begin{tabular}{c|ccccc}
\toprule
maps&\makecell[c]{PPO\\epochs}&\makecell[c]{PPO\\clip}&\makecell[c]{policy loss\\coefficient}&\makecell[c]{lr decay\\strategy}\\
\midrule
5m\_vs\_6m&15&0.05&5&linear\\
8m\_vs\_9m&15&0.1&5&linear \\
10m\_vs\_11m&15&0.1&5&linear\\
27m\_vs\_30m&15&0.1&5&linear \\
6h\_vs\_8z&15&0.05&2&linear \\
MMM&15&0.2&5&linear\\
3s5z&10&0.2&5&linear\\
MMM2&15&0.05&5&exponential\\
3s5z\_vs\_3s6z&10&0.1&2&linear\\
\bottomrule
\end{tabular}
\end{table}

\section{Details of Experimental Results}

\begin{figure*}[htbp]
\centering
\includegraphics[width=1\linewidth]{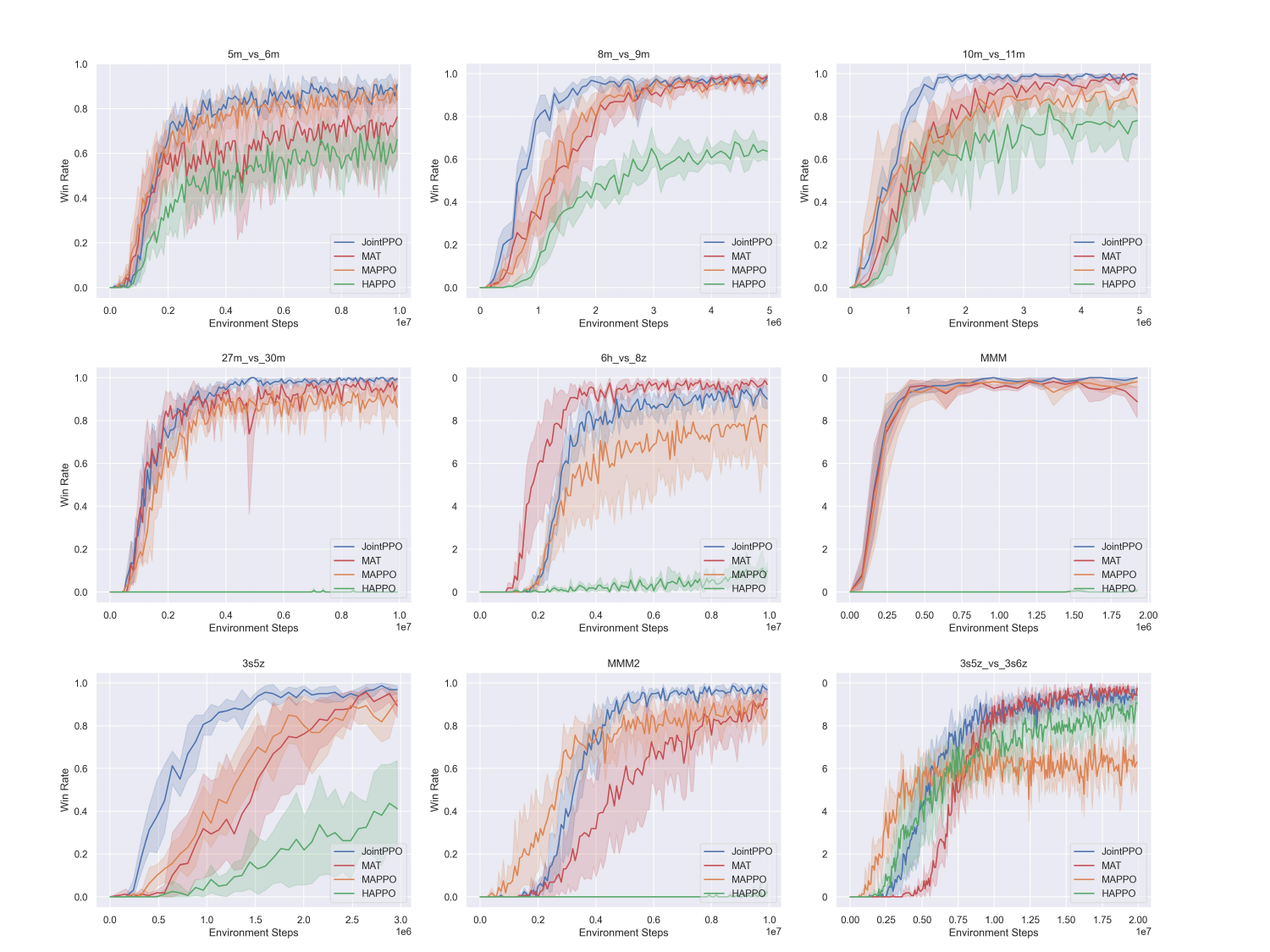}
\caption{Performance comparison on SMAC tasks in terms of win rate.}
\label{fig8}
\end{figure*}

\begin{figure*}[htbp]
\centering
\includegraphics[width=1\linewidth]{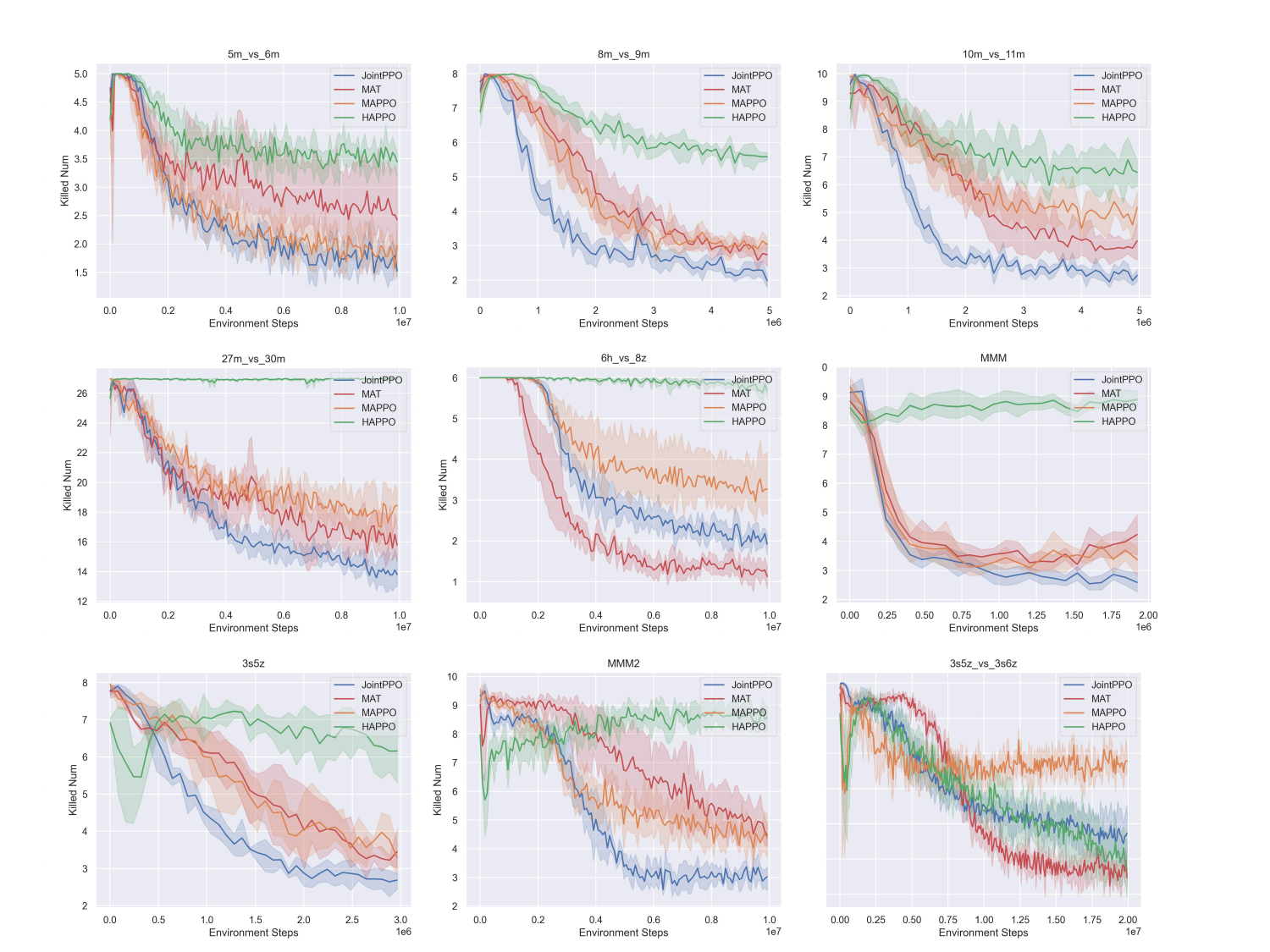}
\caption{Performance comparison on SMAC tasks in terms of the number of killed allies.}
\label{fig9}
\end{figure*}

In this section, we present details of the experiment results, including the training curves of win rates and number of killed allies across all test maps in Figure \ref{fig8} and Figure \ref{fig9}. We also present the detailed results of ablation study on the influence of ppo epoch and clipping parameter. We record the \textbf{final win rate} and \textbf{average win rate} for each set of parameters, seen in Table \ref{tab:ablation1}. The \textbf{final win rate} is the win rate described above which reflects the optimality of the convergence, while the \textbf{average win rate}, here we refer to the win rate averaged over the entire training process from scratch, which reflects the learning rate and the sample efficiency. Both kinds of win rate are averaged over 5 random seeds.

\begin{table}[htbp]
\centering
\caption{PPO Epochs and Clipping Parameter Ablations}
\label{tab:ablation1}
\resizebox{1.0\linewidth}{!}{
\begin{tabular}{|c|c|c|c|c|}
\hline
\diagbox{Epochs}{Clipping} & 0.05 & 0.1 & 0.15 & 0.2 \\
\hline
5 & 90.0 (46.8) & 85.3 (55.6) & 88.1 (64.6) & 73.1 (51.4) \\
\hline
10 & 93.4 (53.6) & 94.1 (66.9) & 92.2 (66.2) & 91.3 (68.1) \\   
\hline
15 & 96.9 (62.7) & 92.8 (67.1) & 90.6 (61.6) & 88.1 (60.7) \\ 
\hline
\end{tabular}
}
\end{table}

\end{document}